\documentclass[aps,pra,twocolumn,amsmath]{revtex4}
\usepackage{graphicx}

\begin{document}

\title{Fault-tolerant linear optics quantum computation by error-detecting quantum state transfer}
\author{Jaeyoon Cho}
\affiliation{Division of Advanced Technology, Korea Research Institute of Standards and Science, Daejeon 305-340, Korea}
\date{\today}
\begin{abstract}
A scheme for linear optical implementation of fault-tolerant quantum computation is proposed, which is based on an error-detecting code. Each computational step is mediated by transfer of quantum information into an ancilla system embedding error-detection capability. Photons are assumed to be subjected to both photon loss and depolarization, and the threshold region of their strengths for scalable quantum computation is obtained, together with the amount of physical resources consumed. Compared to currently known results, the present scheme reduces the resource requirement, while yielding a comparable threshold region.
\end{abstract}
\maketitle

\newcommand{\ket}[1]{\left|{#1}\right>}

\section{Introduction}

One of the main obstacles to implementing a quantum computer using single-photon qubits is the lack of high nonlinearity between individual photons. Thanks to the development of the linear optics quantum computation (LOQC) scheme \cite{klm01}, we now believe that such a problem would be solved in terms of measurement-induced nonlinearity. Recently the demanding requirements of the original LOQC scheme have been significantly reduced by importing the idea of one-way quantum computation \cite{rb01,n04,br05}. In this context, much effort has been devoted to improving the key experimental techniques involved in it, such as generation of single-photon cluster states \cite{wrr05,ksw05,lzg07} and storing single photons coherently in fiber loops for a period of time needed for feed-forward \cite{pjf02,pwt07}.

Although the LOQC approach seems to be quite promising, we are remained with another essential requirement for the practical realization: an ability to cope with inevitable physical noise originated from decoherence or imperfect operations. Fortunately, we are equipped with the ingenious theory of fault-tolerant quantum computation \cite{s96,ab97,klz98,p98,s03,k05}. The main result of it is the threshold theorem which states that a scalable quantum computation can be performed with an arbitrary precision provided the noise strength is below a certain threshold. 

Applying the threshold theorem to the model of one-way quantum computation is not a trivial task. The main reason is that the threshold theorem was originally devised for the quantum circuit model. Recently, there have been reports on the existence of the fault-tolerant threshold for one-way quantum computation \cite{nd05} and also on the estimation of its value \cite{al06,rh07}.

Concerning LOQC, however, it is more difficult to obtain the threshold result, since additional effects due both to the non-deterministic nature of optical two-qubit gates and to photon loss have to be taken into account. In Ref.~\cite{dhn06l}, this issue was addressed and the fault-tolerant threshold was calculated by introducing an error-correction scheme tailored to LOQC. The basic idea was putting a layer bridging the physical level, wherein non-deterministic two-qubit gates are used, and the higher levels, wherein fault-tolerant quantum circuit computation is performed as usual. In the bridging layer, such non-determinism of two-qubit operation is overcome by off-line preparation of ancilla cluster states, exploiting a massive amount of parallel operations and filtering processes. Although it is a remarkable result, the problem is that its resource requirement is extremely demanding, which leaves much room for improvement.

In fact, most of the resources in the above scheme are consumed in the bridging layer, wherein two-qubit gates with a relatively low success probability have to be used so as to embed the error correction into the ancilla state and thus a massive amount of photons are filtered out. Therefore, if this layer is simplified by using a smaller code, the resource consumption might be significantly reduced. This reasoning is the starting point of the present work. In this paper, a fault-tolerant LOQC scheme using an error-detecting code is proposed. The use of an error-detecting code and a consequent error-detecting scheme, which are much simpler than the error-correcting counterparts, results in the reduction of the resource consumption in many orders of magnitude. Remarkably, this is done without the expense of decreasing the tolerable noise level. 

The present work is on the same footing as recent investigations of LOQC based on polarization-entangled single photon qubits, fusion gates, and one-way quantum computation \cite{br05,dhn06a}. The details of the physical model are described in Sec.~\ref{sec:model}. In Sec.~\ref{sec:scheme} a fault-tolerant quantum computation scheme using an error-detecting code is introduced. This scheme is similar in many respects to that of Ref.~\cite{k05} albeit a different error-detecing code is used and modifications are made to tailor it to LOQC. Sec.~\ref{sec:implementation} explains how this fault-tolerant quantum computation scheme is implemented with the resources described in Sec.~\ref{sec:model}. This LOQC scheme is simulated using the numerical method outlined in Sec.~\ref{sec:simul} with the noises described in Sec.~\ref{sec:model} being taken into account, and the numerical results of the error statistics are surveyed in Sec.~\ref{sec:result}. Finally, the conclusion is given in Sec.~\ref{sec:conclusion}.

\section{Physical model\label{sec:model}}

\subsection{Physical resources and elementary operations}

We are supposed to be given a sufficient amount of (a) two-photon polarization-entangled Bell states, (b) linear optical elements such as beam splitters and waveplates, (c) photodetectors resolving photon numbers, and (d) quantum memory gates that are used to store and retrieve single photons without affecting the polarization states. We denote the computational basis states of a single photon qubit, or the eigenstates of Pauli operator $Z$, by $\ket0$ and $\ket1$. 

The underlying idea is using these resources to generate cluster states of single photon qubits and measuring individual photons of them to simulate quantum logic operations. In spite of the intrinsically probabilistic nature, a linear optical two-qubit gate can be used to build up a large-sized cluster state efficiently provided it works in a conclusive manner, i.e., it succeeds only with a probability less than one but whether the operation has succeeded or not can be detected. To build up a large-sized cluster state, one starts with combing resources (a), which are in fact two-qubit cluster states, into larger cluster states using two-qubit gates, and proceeds with combining successfully combined cluster states into more larger ones. This process thus requires a large amount of parallel operations and classical feed-forwards.

The (type-I) fusion gate is well suited for this purpose \cite{br05}. It is implemented by mixing two photons on a polarizing beam splitter and measuring one of the output ports in the basis of $\ket{\pm}=\frac{1}{\sqrt2}(\ket0\pm\ket1)$, the eigenstates of Pauli operator $X$. In case two photons included in cluster states are input to a fusion gate, one and only one photon is detected with a probability $1/2$, in which case, two-photon input states $\ket{00}$ and $\ket{11}$ are projected respectively into single-photon output states $\ket{0}$ and $\ket{1}$, resulting in combining the cluster states into a larger one. In other cases, wherein no or two photons are detected, both input photons are, in effect, measured in the $Z$ basis, resulting in simply removing the two photons from the cluster states.

\subsection{Noise model and assumptions\label{subsec:noise}}

As in the conventional theory of fault-tolerant quantum computation, the whole computational process is assumed to be split into unit time steps. In our case, one time step corresponds to the time taken by one measurement operation (including the ensuing classical feed-forward), which is the dominant time scale involved in the real experiment \cite{pwt07}. Each memory gate, fusion gate, and measurement thus takes one time step, and Bell states are prepared before any desired time steps. At each time step, each photon undergoes two types of noise independently: photon loss and depolarization, whose respective strengths are parameterized by $\gamma$ and $\epsilon$. Each photon is lost independently with probability $\gamma$ after Bell state preparation, before a memory gate, before a fusion gate, and before measurement. Depolarization is simulated by applying randomly chosen Pauli operators ($X$, $Z$, and $Y=XZ$) at relevant {\em locations}: after Bell state preparation and before a fusion gate, one of the 15 possible non-identity Pauli products is applied to the two photons each with probability ${\epsilon}/{15}$; before a memory gate and before measurement, one of the 3 Pauli operators is applied to the photon each with probability ${\epsilon}/{3}$. We also make the conventional assumptions: any pair of photons can be input to a fusion gate at any time without additional overhead; any number of elementary operations can be performed in parallel; processing of classical information takes a sufficiently small time  and is error-free.

\section{Fault-tolerant quantum computation by error-detecting quantum state transfer\label{sec:scheme}}

\subsection{Pauli frame}

Instead of implementing Pauli operators physically, we just keep track of a product of Pauli operators which should have been applied to the state. This  Pauli product is called the Pauli frame \cite{k05,dhn06a,da07}. The Pauli frame is updated after each operation according to the quantum circuit identities. This is possible since we use the Clifford gates that transform Pauli operators into Pauli operators under conjugation. In the same fashion, it is also possible to interpret measurement results with respect to the Pauli frame. In our case, all measurements involved in the error detection are performed in the $X$ basis, thus the measurement value is flipped if the corresponding Pauli frame is $Z$ or $Y$. The use of the Pauli frame and the appropriate measurement basis greatly reduces the amount of feed-forwards imposed in the original idea of one-way quantum computation. It is important because feed-forward is one of the main bottlenecks in LOQC.

\begin{figure}
\includegraphics[width=0.4\textwidth]{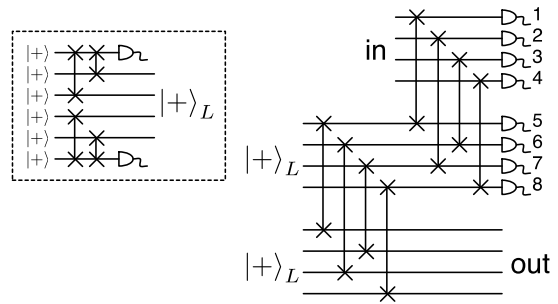}
\caption{\label{fig:edqst}Quantum circuit for error-detecting quantum state transfer. The inset shows a circuit for the generation of $\ket+_L$, the $+1$ eigenstate of the encoded Pauli operator $X_L$.}
\end{figure}

\subsection{Error-detecting code and encoded operations\label{subsec:encoding}}

We use a 4-qubit stabilizer code, which is stabilized by three operators: $\bar{S}_1=X_1X_2X_3X_4$, $\bar{S}_2=Z_1Z_2I_3I_4$, and $\bar{S}_3=I_1I_2Z_3Z_4$, where $X_i$ and $Z_i$ are the Pauli operators and $I_i$ is the identity operator acting on the $i$th qubit. Since the distance of this code is two, it detects one error. The encoded Pauli operators are chosen to be $Z_L=Z_1I_2Z_3I_4$ and $X_L=X_1X_2I_3I_4$. The corresponding encoded eigenstates are then given by $\ket0_L=\frac{1}{\sqrt2}(\ket{0000}+\ket{1111})$, $\ket1_L=\frac{1}{\sqrt2}(\ket{0011}+\ket{1100})$, and $\ket{\pm}_L=\frac12(\ket{00}\pm\ket{11})(\ket{00}\pm\ket{11})$. This choice simplifies the generation of an encoded state $\ket{+}_L$ since it decomposes into two Bell states. Each Bell state is generated by measuring the middle qubit of a linear three-qubit cluster state in the $X$ basis, as shown in the inset of Fig.~\ref{fig:edqst}. This is a transversal operation thanks to the symmetry of Bell states. Note that this process can be interpreted as generating a two-qubit cluster state $\frac{1}{\sqrt2}(\ket0\ket+ + \ket1\ket-)$ and performing an Hadamard operation to the second qubit, where a $\ket+$ state, a CPHASE gate, and an $X$-basis measurement are regarded as constituting an Hadamard operation. At the level of encoded states, this is actually the simplest way of performing a transversal Hadamard operation. The encoded CPHASE operation is performed by four disjoint CPHASE operations as $U(1,5)U(2,7)U(3,6)U(4,8)$, where $U(i,j)$ denotes CPHASE between qubits $i$ and $j$, and two encoded qubits are represented, respectively, by qubits 1 to 4 and 5 to 8. The measurement of the encoded operator ${X}_L$ or ${Z}_L$ is performed by measuring each of the four qubits in the same basis of $X$ or $Z$. Note that these measurements give redundant information. For instance, the value of the ${X}_L$-basis measurement is given either by the measurement of $X_1X_2$ or $X_3X_4$. This redundancy plays an important role in the code concatenation described later. 

\subsection{Error-detecting quantum state transfer}

Fig.~\ref{fig:edqst} depicts a quantum circuit for an encoded memory gate.  Two encoded ancilla qubits are first prepared in an encoded cluster state and then connected to an encoded input qubit through an encoded CPHASE gate. We then measure both the input qubit and the next ancilla qubit in the ${X}_L$ basis. It is easily seen that this process simply transfers the encoded input state to the last encoded ancilla qubit up to the Pauli frame correction \cite{rb01}. An important insight into this process, which we will call the error-detecting quantum state transfer, is that it also embeds the syndrome information into the measurement results. Suppose first that the quantum circuit in Fig.~\ref{fig:edqst} is error-free. If one of the input qubits has an $X$ error, the corresponding value among those of measurements 5 to 8 is flipped, which leads to the parity of the four measurement values being odd. In the same manner, one $Z$ error in the input leads to the parity for the measurement of $X_1X_2X_3X_4$ being odd. One can easily check that a Pauli error at any one location in the whole circuit is also indicated by the parity checks, or possibly by the next round of error-detecting quantum state transfer. The idea outlined here can be easily extended in such a way that an encoded operation is also embedded into the circuit by preparing a different ancilla state. The details will be described later.

\subsection{Code concatenation}

We concatenate the encoding described above hierarchically in such a way that four qubits at each level constitute one qubit at the next higher level. This hierarchy of concatenation ranges from level 0, at which a qubit is encoded in one photon, to level $l_d$, at which a qubit is encoded in $4^{l_d}$ photons. One objective of the present work is to obtain the statistics of errors occurring in each operation at each level. In the conventional theory using an error-correcting code, they are described in terms of the error rate, the probability that the operation yields a logically incorrect output, i.e., an output with errors that can not be corrected even by noiseless error correction (in case the output has correctable errors, they are counted in obtaining the error rate of the next operation). Error rates obtained in this way for every individual operations at a certain level are used as parameters for obtaining error rates at the next higher level. For instance, let us consider a case of one-error-correcting code and denote the error rate at level 0 by $p$. At level 1, the error rates behave as $p^2$ since every events occurring with probability of order $p$ are those having an error at a single location, which can be corrected. In the same fashion, the error rates at level $l$ behave as $p^{2^l}$. They should decrease rapidly as long as the initial error rate $p$ is sufficiently small; more precisely, as long as $p$ is less than the fault-tolerant threshold.

In our case, it is necessary to distinguish between two types of errors, namely, {\em located} and {\em unlocated} errors. If a set of Pauli errors and the locations at which they occurred is such that the embedded error-detecting process detects it, it is counted as a located error. Accordingly, the located error rate is defined as the probability of occurrence of such errors. On the other hand, if the set contains Pauli errors and locations which are not detected by the embedded process while causing a logically incorrect output, it is counted as an unlocated error (detectable errors remaining in the output are counted in obtaining the error rates of the next operation). The unlocated error rate is defined rather differently as the probability of occurrence of an unlocated error given no located error occurs. The reason it is defined as such a conditional probability is that ancilla states having located errors are all discarded from the computation. Suppose that the located and unlocated error rates at a certain level are respectively of order $q$ and $p$. Those at the next higher level then behave respectively as $O(q)+O(p)$ and $O(p^2)$. The problem is that, as this estimation indicates, error detection does not decrease located error rates. Consequently, the previous scenario in which error correction is used does not work in our case. We overcome this difficulty by exploiting the redundancy of encoded Pauli measurements, explained in Sec.~\ref{subsec:encoding}. For example, if a locate error occurs during the CPHASE operation corresponding to measurements 2 and 7 in Fig.~\ref{fig:edqst}, instead of discarding the operation, we use the results of measurements 3, 4 and 5, 6 to obtain the values of the two $X_L$-basis measurements (of course, in the ancilla preparation step, no located error is allowed). By using this method, the located and unlocated error rates of the encoded operation behave respectively as $O(p)+O(q^2)$ and $O(qp)+O(p^2)$. For sufficiently small $q$ and $p$, these error rates decay exponentially as the level of encoding gets higher, as in case of error correction. We will use both of the above decoding methods, and call the former the {\em strong-detection mode} and the latter the {\em weak-detection mode}. For brevity, we will also use the terms {\em strongly-detected} and {\em weakly-detected} operations. Note that although we classify each operation further into two types according to how the syndrome information is decoded, there is no difference in their physical implementation. The basic strategy is as follows: In the ancilla preparation stage, we mainly use strongly-detected operations to reduce error rates as much as possible. If any located error occurs, the ancilla state is discarded and the preparation is restarted. Once an ancilla state is successfully generated, we use weakly-detected CPHASE operations to connect it to an input state, in which way we reduce the chance of destroying the main computation.

\subsection{Universal set of quantum gates}

The universal quantum computation is guaranteed by bringing in the preparation of state $\ket{\pi/8}=\cos(\pi/8)\ket0+\sin(\pi/8)\ket1$ which allows the implementation of the $\pi/8$ gate $T=\exp(-i\frac{\pi}{8}Z)$. In order to prepare the $\ket{\pi/8}$ state, we first prepare a Bell state at the topmost level. We then measure one of the two qubits in the basis of $\{\ket{\pi/8},\ket{5\pi/8}\}$, where $\ket{5\pi/8}=ZX\ket{\pi/8}$ is the state orthogonal to the $\ket{\pi/8}$ state. This measurement is done by measuring the corresponding lower-level qubits, respectively, in the bases of $Z$, $Z$, $X$, and $\{\ket{\pi/8},\ket{5\pi/8}\}$, the last of which is, in turn, done by measuring the lower-level qubits in the same way. As a result, the other qubit at the topmost level is remained in the $\ket{\pi/8}$ state up to the Pauli frame correction. Note that errors occurring during the measurement could introduce an error at the topmost level, but they do not destroy the encoding structure. We can thus purify multiple copies of noisy $\ket{\pi/8}$ states at the topmost level \cite{k05,k04,bk05}. The error rates of this preparation is expected to be determined dominantly by those of measurements at low levels, which is confirmed by numerical simulation. The unlocated error rate of the preparation of a $\ket{\pi/8}$ state is found to be well within the bounds for successful purification.

\section{Linear optical implementation\label{sec:implementation}}

\subsection{Level-0 encoding}

Quantum state transfer at the bottom level relies on microclusters and parallel fusion as outlined in Ref.~\cite{dhn06a}. Quantum information is stored in the center node of a star-shaped cluster state, called a microcluster. In order to transfer the state of one center node, say qubit 1, to another center node, say qubit 2, we apply fusion gates pairwise between the dangling nodes in parallel, and measure qubit 1 and all successfully fused nodes in the $X$ basis. In case one or more fusion gates succeed, the state of qubit 1 is transferred to qubit 2 up to the Pauli frame correction. If the number of fused nodes is two or more, the ensuing measurements of them should give the same value, if not affected by noises. If it is not the case, a majority vote is taken over the values, and if it is not allowed, i.e., half of them are $+1$ and the rest are $-1$, we conclude qubit 2 has a located error. We also come to the same conclusion if any of the involved operations indicates a photon loss. If we ignore noise, the success probability of the parallel fusion increases exponentially as the number of dangling nodes increases. More dangling nodes, however, introduce more noise and thus decrease the success probability asymptotically. In this paper, the number of dangling nodes of each microcluster is fixed as 4. Note that, throughout this paper, we will always regard a fusion gate and the ensuing $X$-basis measurement as being performed in one time step, since it can be done by measuring both the output photons of the polarizing beam splitter in the $X$ basis. The noise model for this combined operation will just follow the rules described in Sec.~\ref{subsec:noise} as if it is performed in two time steps.

\subsection{Level-1 encoding}

\begin{figure}
\includegraphics[width=0.45\textwidth]{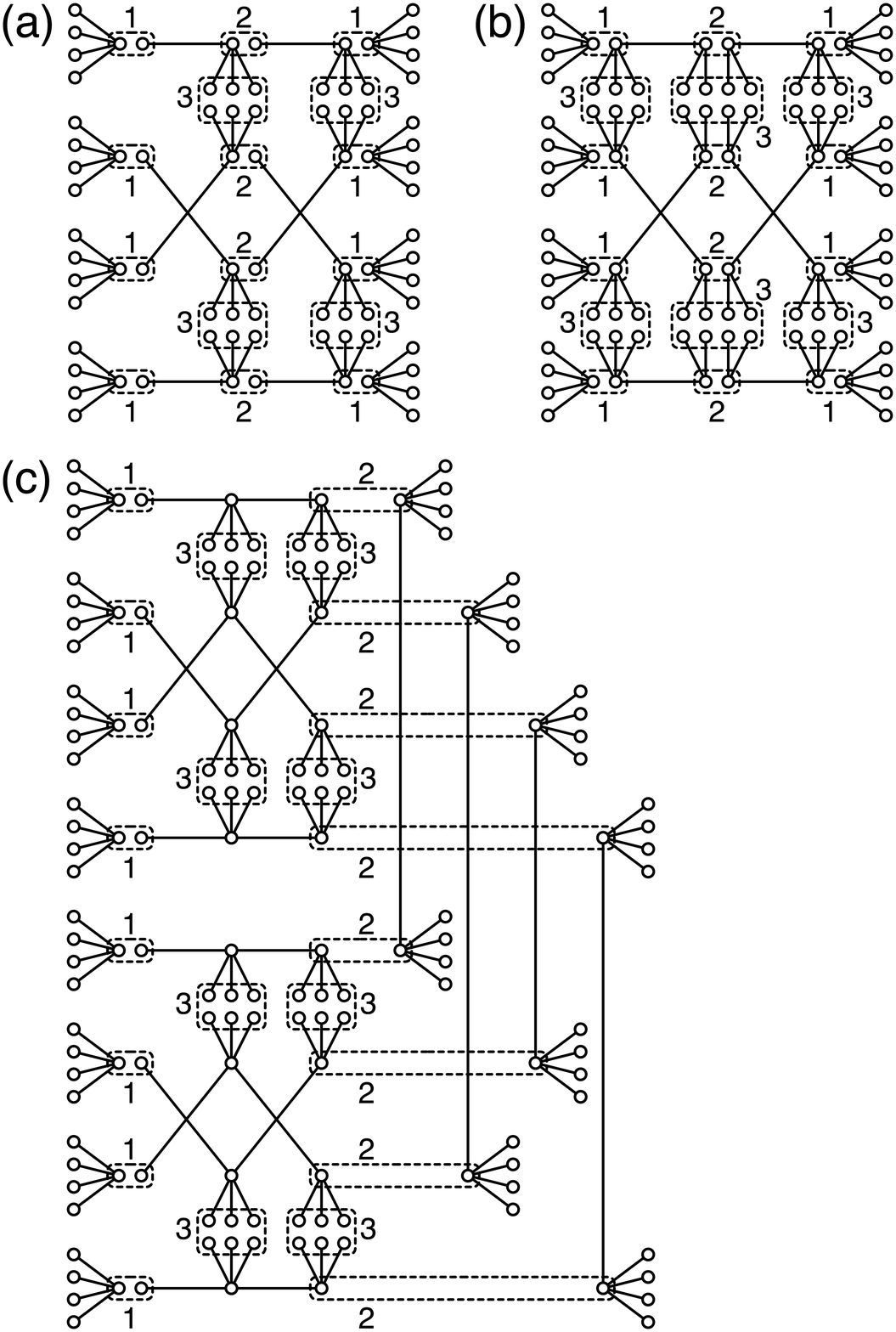}
\caption{\label{fig:level1}Cluster states for the level-1 encoding: (a) a memory gate, (b) a Bell-state, and (c) a CPHASE gate. Fusion gates are applied in the order indicated by the numbers.}
\end{figure}

Fig.~\ref{fig:level1}(a) depicts the cluster state to simulate a level-1 memory gate. Each circle represents a single-photon qubit. Fusion gates are applied in the order indicated by the numbers. The resulting cluster state is then composed of three columns, each representing a level-1 qubit, and the dangling nodes at both sides (leftmost nodes for input and rightmost nodes for output). It is easily seen that the quantum circuit simulated by this cluster state is equivalent to that of error-detecting quantum state transfer shown in Fig.~\ref{fig:edqst}. All qubits in the three columns are measured in the $X$ basis. Note that these measurements can be done even before an input state is fused (an analogous idea is used for the telecorrector introduced in Ref.~\cite{dhn06a}). Actually, each measurement is performed together with a preceding fusion gate in one time step as explained earlier. If any of the fusion gates or the measurements indicates a located error, this preparation stage is restarted. Otherwise, the resulting state is accepted and finally fused with a level-1 input qubit on the left side. Note that the measurements in the rightmost column do not give information about the encoded state, since it is transferred to the rightmost dangling nodes. 

This idea can be easily extended to other operations. For an $X$-basis measurement, we construct the same cluster state without the rightmost dangling nodes. Note that in this case the measurement parity of the rightmost column is not altered by the Pauli frame correction performed after fusing an input qubit. If this measurement parity is found to be odd, it means the ancilla state has an error and we thus restart the preparation stage. A CPHASE operation is performed in a similar way by constructing a cluster state composed of six columns of qubits, say columns 1 to 6, as shown in Fig.~\ref{fig:level1}(c). In this case, columns 1 and 6 have dangling nodes for the input, while columns 3 and 4 have dangling nodes for the output. In case a CPHASE gate is followed by measurements, they are also embedded into the preparation stage. For example, when the first output qubit of a CPHASE gate should be measured, we construct a cluster state composed of six columns with dangling nodes attached to columns 1, 4, and 6, and the measurement parity of column 3 is used to filter out noisy ancilla states. Another useful operation is the Bell-state preparation shown in Fig.~\ref{fig:level1}(b), which are used for the preparation of a level-2 $\ket+$ state. The resulting state is accepted unless any photon loss is detected or the measurement parity of the middle column is found to be odd.

Remarkably, once appropriate cluster states are generated (with all qubits being measured except at the dangling nodes), every level-1 gate operations are performed in one time step by fusing and measuring dangling nodes in parallel. Level-1 gates can be thus treated as if they are unit gates occupying one time step in a level-2 quantum circuit with a particular located and unlocated error rate. Therefore, once the error rates of individual level-1 gates are calculated for given level-0 noise rates $\gamma$ and $\epsilon$, the problem reduces to that of determining whether the set of level-1 error rates is within the fault-tolerant threshold of the conventional quantum circuit model.

\subsection{Second and higher levels of encoding\label{subsec:highlevel}}

\begin{figure}
\includegraphics[width=0.47\textwidth]{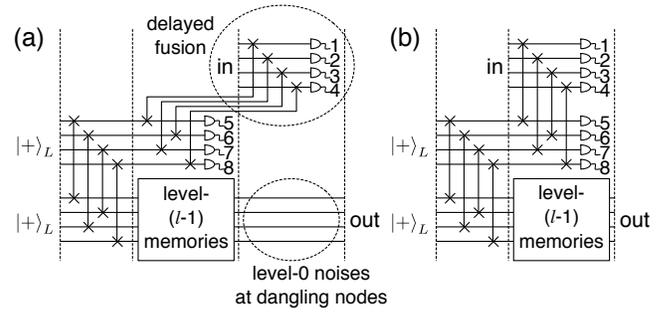}
\caption{\label{fig:memory2}Quantum circuit for a level-$l$ memory gate (a) for $2\le l \le l_c$ and (b) for $l> l_c$. Each column represents one time step.}
\end{figure}

\begin{figure}
\includegraphics[width=0.47\textwidth]{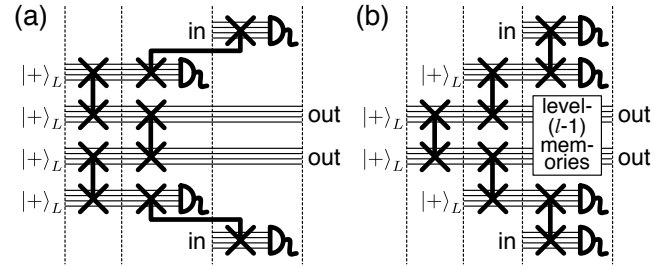}
\caption{\label{fig:cphase2}Quantum circuit for a level-$l$ CPHASE gate (a) for $2\le l\le l_c$ and (b) for $l>l_c$.  Each gate depicted with thick lines is an abbreviation of the corresponding four level-$(l-1)$ gates. Each column represents one time step.}
\end{figure}

From the second level of encoding, each gate operation is performed by directly implementing the quantum circuit shown in Fig.~\ref{fig:edqst} or its variant. A normal way of doing this would be to arrange involved lower-level gates consecutively along the time axis in a way as depicted in Fig.~\ref{fig:memory2}(b) for a memory gate or Fig.~\ref{fig:cphase2}(b) for a CPHASE gate. This method, however, requires error rates lower than can be achieved by our level-1 gates. This is the case even if the physical noises are neglected ($\gamma=\epsilon=0$), owing to the probabilistic nature of the fusion gate. In such a case, the level-0 state transfer fails with probability $0.5^4=0.0625$, as we use four dangling nodes per qubit, and this gives the level-1 located error rate of about $0.0215$. This value is higher than the located error rate required for the memory gate shown in Fig.~\ref{fig:memory2}(b) to work correctly, which is about 0.0210 even if the unlocated errors are all neglected. 

We overcome this difficulty by introducing {\em delayed fusion}, which exploits the fact that two qubits can be input to a CPHASE gate at different time steps. For instance, we modify the quantum circuit for a memory gate as shown in Fig.~\ref{fig:memory2}(a), and in the same fashion a CPHASE gate as shown in Fig.~\ref{fig:cphase2}(a). For a memory gate, we first implement the ancilla part up to the half of the final gates (CPHASE plus measurement), corresponding to measurements 5 to 8, while the input dangling nodes at the other half are left untouched. Only when this preparation stage is successful, an input qubit is finally fused, during which the output dangling nodes are left untouched. Compared to the circuit in Fig.~\ref{fig:memory2}(b), this method puts more larger parts into the preparation stage so that located errors on them can be also filtered out. As a result, the located error rate of the encoded gate is much reduced. However, this also introduces idle time steps of dangling nodes causing additional level-$0$ memory noises. Moreover, these noises accumulate as the level of encoding gets higher. That is, a level-2 gate uses level-1 gates delayed one time step, but a level-3 gate uses level-2 gates delayed one time step which in turn use level-1 gates delayed two time steps. After the accumulated noise exceeds a certain bound, additional levels of concatenation will make things worse rather than better. Therefore, we use this method only until we reach the $l_c$th level of encoding in which we get sufficiently low error rates, and from the $(l_c+1)$th level, we proceed with the implementation normally without the delayed fusion as in Fig.~\ref{fig:memory2}(b) and Fig.~\ref{fig:cphase2}(b).

\section{Numerical simulation\label{sec:simul}}

Error rates are calculated numerically using the Monte Carlo simulation \cite{dhn06a}. For each iteration of the simulation, randomly chosen errors are assigned to every involved gates according to the error rates, and the resulting error in the encoded output qubit is identified. In so doing, we do not have to care about the full state of the involved qubits, since our concern is the error statistics that can be completely predicted just by recording the errors together with the locations at which they occurred. The details can be best illustrated by considering several examples.

\begin{figure}
\includegraphics[width=0.3\textwidth]{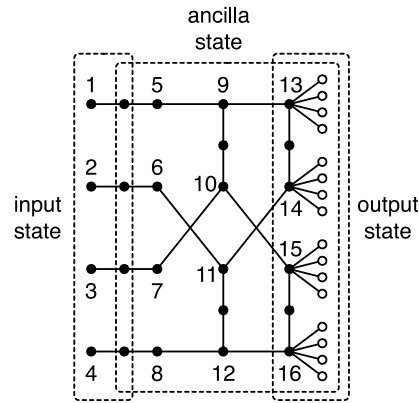}
\caption{\label{fig:memory1}Cluster state equivalent to that generated during the level-1 memory operation. Every qubits represented by filled circles are measured in the $X$ basis.}
\end{figure}

Let us first consider the strongly-detected level-1 memory gate. It is implemented by preparing an ancilla state as shown in Fig.~\ref{fig:level1}(a), which involves sixteen star-shaped five-qubit cluster states. Each of them is generated by fusing middle qubits of three-qubit linear cluster states, each of which is in turn generated by fusing two-qubit cluster states. Note that in doing this bottom-up construction, all involved states are cluster states (with corresponding Pauli frames). Any cluster state with a Pauli error (i.e., an error given by a product of Pauli operators) can be transformed to a noiseless state with an incorrect Pauli frame having an error composed of only $Z$ operators. It is easily seen by observing a quantum circuit representing such a noisy cluster state composed of $\ket{+}$ states, CPHASE gates, and Pauli gates representing the Pauli frame and the Pauli errors. From the quantum circuit identities, the order of a CPHASE and an $X$ gate can be exchanged by adding a $Z$ gate, whereas the order of a CHASE and a $Z$ gate can be freely exchanged. By using these properties, every $X$ errors can be put next to the $\ket{+}$ states, and using $X\ket+=\ket+$, we are remained with an error composed of only $Z$ operators. Note that a phase factor $-1$ arising in exchanging the order of $X$ and $Z$ is a global phase which has no practical meaning. To simulate a noisy fusion gate, we first apply Pauli errors randomly according to the rule in Sec.~\ref{subsec:noise}, transform all errors to $Z$ errors in Pauli frames, and apply a noiseless fusion gate. Considering only $Z$ errors in Pauli frames, a noiseless fusion gate works in such a way that the output qubit has a $Z$ error only in case either of input qubits has a $Z$ error while the other has no error. Noisy measurements are simulated in a similar way: since all qubits are measured in the $X$ basis, a measurement yields an incorrect result only in case the qubit has a $Z$ error in the Pauli frame. Once a noisy cluster state for a level-1 memory is generated in this way and if no located error is detected in it, it is fused with an input state having 4 qubits and 16 dangling nodes attached to them, and we proceed as follows. 

(1) If any of the four parallel fusions between the input and the ancilla indicates a located error, this memory gate is counted as having a located error. Otherwise, we proceed to the next step.

(2) The state is now equivalent to the cluster state shown in Fig.~\ref{fig:memory1} with every qubits represented by filled circles being measured in the $X$ basis. The next step is to simulate the propagation of errors from the input to the ancilla state. For example, if qubit 1 in Fig.~\ref{fig:memory1} has a $Z$ error, it is propagated to qubit 5. Although such an error also induce an $X$ error at the fused node between qubits 1 and 5, it has no effect as the node is measured in the $X$ basis. In the same manner, if the fused node between qubits 2 and 6 has a $Z$ error, it is propagated to qubit 11.

(3) The syndrome measurement for qubits 5 to 8 is examined. If the number of $Z$ errors among them is one or three, the error is detected, thus this memory gate is counted as having a located error. On the other hand, if the number of $Z$ errors is two and if one of them exists at either qubit 5 or 6 while the other at either qubit 7 or 8, the error flips the value of this $X_L$-basis measurement, resulting in incorrect update of the Pauli frame. This is reflected by applying $Z$ errors to qubits 13 and 15.

(4) The syndrome measurement for qubits 9 to 12 is examined in the same way as in step (3). If the error is detected, this memory gate is counted as having a located error. On the other hand, if the error flips the value of this $X_L$-basis measurement, instead of pretending the incorrect Pauli update as in step (3), we just conclude this memory gate has an unlocated $X$ error (note that we have performed a transformation so that there are no more $X$ errors in the ancilla state). We do not pretend the incorrect Pauli frame update, since it is caused by the error that has been already counted.

(5) The $Z$ errors at qubits 13 to 16 are examined. If the number of $Z$ errors is two and if they exist respectively at qubits 13, 14 and qubits 15, 16, we conclude this memory gate has an unlocated $Z$ error, and this unlocated error is removed from the state, since it has been counted.

(6) If this memory gate has both unlocated $X$ and $Z$ errors, the error is counted as an unlocated $Y$ error instead. The remaining level-1 output qubit, i.e., qubits 13 to 16 and the attached dangling nodes, is used as an input state for the next round of the simulation. 

By following these steps, one sample of the Monte Carlo simulation is yielded. A remaining problem is, however, that the error rates depend on the errors in the input state as well. A standard treatment is applying memory gates successively to a noiseless initial state until the transient effect washes out, and taking the next memory gate to quantify the error statistics \cite{dhn06a,s03}. Actually, such a transient effect rapidly washes out in our case as well. It can be shown numerically that after one memory operation, each successive memory gate exhibits the same error rates. We thus repeat applying a level-1 memory gate to a noiseless input state until the gate yields no located error and take the output of it as an input for another level-1 memory gate to obtain one sample of the simulation. In fact, at the second level of encoding, level-1 memory gates are used not after memory gates but after CPHASE gates as shown in Fig.~\ref{fig:memory2}, but that does not make substantial changes in the error rates.

The error rates of the level-1 Bell-state preparation shown in Fig.~\ref{fig:level1}(b) are also obtained in the same manner. In this case, we use the fact that a Bell state is an eigenstate of $X_1X_2$ and $Z_1Z_2$ to transform all unlocated errors into single-qubit errors. For example, $X_1Z_2$, $Y_1Z_2$, and $Z_1$ errors are respectively equivalent to $Y_2$, $X_2$, and $Z_2$ errors, and so forth. Concerning the level-1 CPHASE gate shown in Fig.~\ref{fig:level1}(c), we simplify the simulation by using a numerically confirmed fact that the errors occur as if level-1 memory errors independently affect two input qubits and an error-free level-1 CPHASE gate follows, making the errors propagate accordingly in such a way that an $X$ error in one qubit induces a $Z$ error in the other qubit. It is also confirmed that the same property holds for all other level-1 gates embedding a CPHASE gate. For example, to simulated a level-1 CPHASE-plus-measurement gate included in the circuit of Fig.~\ref{fig:memory2}(a), we apply the errors of the level-1 measurement and the one-step-delayed level-1 measurement independently to each qubit according to the corresponding error rates, and apply an error-free CPHASE-plus-measurement gate afterward, making the errors propagate accordingly. Note that an $n$-step-delayed measuremet is always performed to an $n$-step-delayed qubit for any $n$, which has to be correctly reflected in the simulation by applying additional level-0 noises to the dangling nodes. To sum up, we just obtain the single-qubit error rates for the Bell-state preparation, memory, measurement, and each $n$-step-delayed measurement, and in case a CPHASE operation is involved, the error rates are obtained by using the single-qubit error rates and the error propagation rule. Note also that, in Fig.~\ref{fig:memory2} or Fig.~\ref{fig:cphase2}, the CPHASE gates next to the Bell-state preparation exhibit different error rates since the inputs are not the outputs of memory gates. This does not arise as a problem since the error rates of them are in fact smaller than those we use, thus we will not underestimate the error rates. 

From the second level of encoding, obtaining the error rates is simpler and consumes less time because we now simulate the errors in the conventional quantum circuit model having a much smaller number of error locations. The underlying method is the same as that used in the level-$1$ simluations: we obtain the single-qubit error rates and use the error propagation rule to obtain the error rates of two-qubit gates. 

\section{Fault-tolerant threshold and resource consumption\label{sec:result}}

The located and unlocated error rates of individual operations are calculated starting from a given set of parameters $\{\gamma,\epsilon\}$, and when the error rates asymptotically go to zero as the level of encoding gets higher, the given set is said to be within the threshold region. In our case, finding out such an asymptotic behavior involves exhaustive calculations since we have to obtain the error rates of every $n$-step-delayed measurements and moreover the asymptotic behavior depends on which level is chosen as $l_c$ [see Sec.~\ref{subsec:highlevel}]. For simplicity, we just fix $l_c=5$ and define one particular {\em sufficient} condition as our fault-tolerance criterion: we say the set $\{\gamma,\epsilon\}$ is within the threshold region if the error rates of the level-$5$ operations satisfy the following conditions:
\begin{align}
\max\{Q^M,Q^S\} & \le 10^{-2},\\ 
\max\{P^B_X,P^B_Z,100P^B_Y \} & \le 10^{-6},\\ 
\max\{P^M_X,P^M_Z,100P^M_Y\} & \le 10^{-4},\\
\max\{P^S_X,P^S_Z,100P^S_Y\} & \le 10^{-4}, 
\end{align}
where $Q$ and $P$ denote, respectively, located and unlocated error rates, subscripts $X$, $Y$, and $Z$ denote the types of the unlocated errors, and superscripts $B$, $M$, and $S$ denote, respectively, Bell-state preparation, memory, and measurement. The required error rates for unlocated $Y$ errors are chosen as smaller values because they are actually found to be much smaller than those of the unlocated $X$ and $Z$ errors. Provided these conditions are met, the error rates at the higher levels can be reduced asymptotically to zero using the weak-detection mode without the need of delayed fusion, as shown in Fig.~\ref{fig:memory2}(b) and Fig.~\ref{fig:cphase2}(b). Note again that these conditions are not necessary ones for the fault tolerance. They are chosen just for convenience so as to prevent exhaustive numerical works.

The error rates and the resource consumption also vary according to how the strong- and the weak-detection modes are incorporated in code concatenation. Here, we consider two methods. In the first method, the ancilla part of a level-$l$ operation is implemented with strongly-detected level-$(l-1)$ operations, while the remaining part is implemented with weakly-detected level-$(l-1)$ operations. The second method is the same as the first method except the ancilla parts of level-2 operations are implemented with weakly-detected level-1 operations. Compared to the first method, this method needs less resources for the same level of concatenation, but leads to higher error rates. 

\begin{figure}
\includegraphics[width=0.35\textwidth]{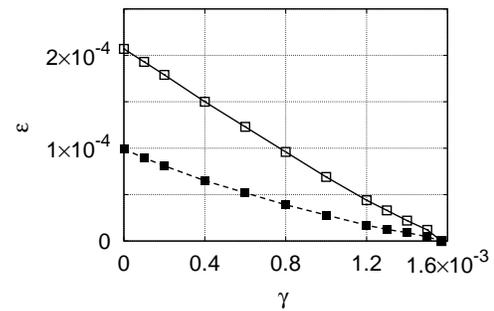}
\caption{\label{fig:threshold}Threshold region for scalable quantum computation using method 1 (solid curve) and method 2 (dotted curve), where $\gamma$ and $\epsilon$ denote, respectively, the probabilities of photon loss and depolarization per operation.}
\end{figure}

The threshold regions for the first method (solid curve) and the second method (dotted curve) are shown in Fig.~\ref{fig:threshold}. To obtain these curves, we first determine the threshold value of $\gamma$ with fixing $\epsilon=0$, which is found to be $1.57\times10^{-3}$. This does not require much computational work, since we have to take into account only photon losses at the dangling nodes of level-0 cluster states and located error rates at higher levels can be easily derived from those of the level-0 operations. In this case, the unlocated error rate is zero. Once the range of $\gamma$ is determined, we start guessing the threshold value of $\epsilon$ with fixing $\gamma$ as particular values, corresponding to dots in Fig.~\ref{fig:threshold}. Near the threshold value, $\epsilon$ is varied by a small increment of $\Delta\epsilon=10^{-6}$, to which extent the threshold value is determined. The number of samples for each Monte Carlo simulation is chosen as $10^7$ at the level-0 encoding and $10^8$ at higher levels. Note that the results in Fig.~\ref{fig:threshold} are comparable to those in Ref.~\cite{dhn06a} using error-correcting codes.

\begin{figure}
\includegraphics[width=0.45\textwidth]{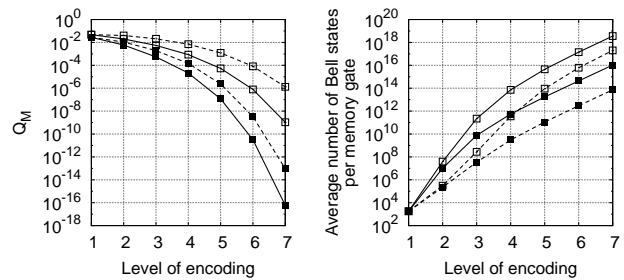}
\caption{\label{fig:resource}Located error rate $Q_M$ of a memory gate (left) and average number of two-photon Bell states consumed for it (right) with respect to the level of encoding.}
\end{figure}

In order to compare the performance further, we plot in Fig.~\ref{fig:resource} the located error rates of a memory gate (left) and the resource consumption for it in terms of the average number of two-photon Bell states used (right) with respect to the level of encoding for the first method (solid curve) and the second method (dotted curve) with two sets of parameters: $\gamma=10\epsilon=4\times10^{-4}$ (unfilled squares) and $\gamma=10\epsilon=10^{-4}$ (filled squares). In these simulations, $l_c$ has been chosen larger than 7, that is, we used delayed measurements at every levels simulated. Note that the curves are obtained with the help of analytic calculations, which are allowed at high levels of encoding wherein located error rates dominate all unlocated error rates and the unlocated error rates of weakly-detected operations dominate those of strongly-detected operations. In such a limiting case, only errors in delayed measurements make dominant contributions to the error rates at the next higher level, since the weak-detection mode is used only for delayed measurements. Consequently, we can easily calculate error rates by taking into account only four error locations associated with them. As located error rates dominate unlocated error rates, they indicate how many operations can be performed reliably. Compared to the results of Ref.~\cite{dhn06a}, our scheme is found to require less resources in general. In case of $\gamma=10\epsilon=4\times10^{-4}$, we obtain the error rate of $10^{-9}$ using about $4\times 10^{18}$ Bell pairs (the first method), while the scheme in Ref.~\cite{dhn06a} requires about $10^{23}$ Bell pairs. The required resources are drastically reduced as the noise strength decreases: in case of $\gamma=10\epsilon=10^{-4}$, the same error rate is attained using about $10^{13}$ Bell pairs (the second method).

\section{Conclusion\label{sec:conclusion}}

In summary, a new scheme for fault-tolerant linear optics quantum computation based on an error-detecting code has been introduced and analyzed with most of the prominent obstacles encountered in linear optics quantum computation, such as non-deterministic two-qubit operation and photon loss as well as imperfect gate operations and decoherence, being taken into account. In order to deal with them, various established techniques and some new ideas were incorporated. The basic building block is the error-detecting quantum state transfer. At the level of encoded qubits, its role is simply transferring quantum information through a linear cluster state, but useful insight is that the intermediate measurements also give the syndrome information for detecting errors. The error-detecting code and encoded operations were concatenated using two types of decoding methods, namely the strong- and weak-detection modes, and introducing delayed fusion operation. It was shown that the rate of error per gate operation can be substantially lowered by the concatenation and both the fault-tolerant region of initial noise rates and the resource consumption were calculated numerically. The numerical results indicate that the resource consumption is decreased in many orders of magnitude compared to the known results. Although the results exhibit great improvements, the resource requirement is still demanding and far from practical implementation. It might be further reduced by importing other techniques such as purification or error correction. In order to make the reduction to an extent that allows practical linear optics quantum computation, however, it seems that new ideas will be required to substantially overcome the non-deterministic nature of two-qubit operation.

\begin{acknowledgments}
This research was supported by the ``Single Quantum-Based Metrology in Nanoscale'' project of the Korea Research Institute of Standards and Science.
\end{acknowledgments}

\bibliography{a}

\end{document}